\documentstyle[aps,twocolumn]{revtex}

\input epsf

\def\prb{Phys. Rev. B}
\def\prl{Phys. Rev. Lett.}

\def\LNiO{La$_2$NiO$_{4+\delta}$}
\def\LSCO{La$_{2-x}$Sr$_x$CuO$_4$}

\def\LNSCO{La$_{1.6-x}$Nd$_{0.4}$Sr$_x$CuO$_{4}$}

\begin{document}
\twocolumn[
\hsize\textwidth\columnwidth\hsize\csname @twocolumnfalse\endcsname
\title{\vspace{-40pt}{\normalsize\null\hspace{5.5in}
} \vspace{30pt}\\
Landau Theory of Stripe Phases in Cuprates and Nickelates}

\author{Oron Zachar ,  S.~A.~Kivelson}
\address
{Dept. of Physics,
UCLA
Los Angeles, CA  90095-1547}
\author{V.~J.~Emery}
\address{
Dept. of Physics,
Brookhaven National Laboratory,
Upton, NY  11973-5000}
\date{\today}
\maketitle 

\begin{abstract}

We consider a Landau theory of coupled charge and spin-density wave 
order parameters as a simple model for the ordering that
has been observed experimentally in the La$_2$NiO$_4$ and
La$_2$CuO$_4$ families of doped antiferromagnets.  The period of the 
charge-density wave is generically half that of the spin-density wave,
or equivalently the charges form antiphase domain walls in
the antiferromagnetic order.  A sharp distinction exists between the case 
in which the 
ordering is primarily charge driven (which produces a sequence of
transitions  in qualitative agreement
with experiment) or spin driven (which does not).  
We also find that stripes with 
non-collinear spin order ({\it i.e.} spiral phases) are possible in
a region of the phase diagram where the transition is spin driven;
the spiral is circular only when there is no charge order, and is otherwise
elliptical with an eccentricity proportional to the magnitude of the
charge order. 
 
\end{abstract}


]

Experiments on the doped lanthanum nickelate \cite{NiO} and lanthanum cuprate 
\cite{NdCuO,CuO} families of materials have established that
topological stripe phases are a prominent feature of doped antiferromagnets.
In the low-temperature stripe phase, the doped holes are concentrated in 
periodic walls which are simultaneously discommensurations in the N\'eel order. 
In this paper, the phase diagram of coupled charge density wave (CDW) and 
spin density wave (SDW) order parameters will be constructed from the Landau 
theory of phase transitions\cite{landau}. 
By associating the experimental observations with 
{\it distinct} regions of the global phase diagram it is possible to gain
insight into the microscopic mechanism of stripe formation and the nature of
the spin ordering.
It will be shown that the experiments are consistent with the suggestion that
stripes are produced by frustrated phase separation
\cite{ps,ps-stripes,spherical} and not by a Fermi surface instability
\cite{HF-stripes}. 

The analysis also
addresses the existence of spiral magnetic order found in some theoretical 
studies of doped antiferromagnets\cite{shraiman}. A circular spiral
phase (in which the magnitude of the ordered moment is a constant)
is only possible if there is no
accompanying charge order;  
the coupling of the spin and charge order
generally produces an elliptical spiral phase, 
with the  eccentricity of the ellipse
proportional to the magnitude of the charge order parameter. 
In the region of the phase diagram which we associate with all known
experiments, spiral phases are a remote possibility. 
They may appear only as a third transition (yet undetected) at very
low temperatures.

It will be assumed that the Landau free energy function depends on only
the fundamental Fourier components of the SDW and CDW order parameters. 
Of course, as always, higher harmonics
will appear as the magnitude of the order increases, but this has
no effect on the nature of the phase diagram.  
So as to focus on the situation of immediate experimental relevance, we will
consider the quasi-two dimensional case in which the ordering vectors
lie in a plane (and, ultimately, in a line), but the generalization to
other geometries is straightforward.  (We shall return to the
issue of the fluctuation effects
peculiar to quasi-two-dimensional systems at the end.)
Also, we consider the case in which the crystal has
the symmetry of a square lattice, and the ordering vectors lie along
a symmetry direction, so that there are only two inequivalent directions
of the ordering vector (along the diagonals in the case of the lanthanum 
nickelates, and along the vertical and horizontal directions for the
lanthanum cuprates). This will allow us to extract the essential physics in
a relatively simple way. The generalizations to the various space groups 
of the structures of any given material (which in principle should be used) 
and to allow different ordering vectors is straightforward.

With these restrictions, the stripe order can be
described by the two complex scalars, $\rho_{\vec{k}}$ and
$\rho_{\vec{k}^{\prime}}$ and the complex spin
vectors ${\bf S}_{\vec{q}}$ and  ${\bf S}_{\vec{q}^{\prime}}$, 
corresponding to charge order and spin order respectively.  Here,
the vectors (${\vec k}$, $\vec{q}$) are related to 
($\vec{k}^{\prime}$, $\vec{q}^{\prime}$) by rotation through $\pi/2$, and 
$\vec q$ is measured relative to the magnetic ordering vector $\vec Q$ of the
undoped system. 
(It is assumed that $\vec Q$ is unique, which requires that 
$\vec Q \equiv -\vec Q$, {\it i.e.} that $2\vec Q$ must be equal to a 
reciprocal lattice vector.) 

The most general free energy up to fourth order is constructed by including 
all terms allowed by symmetry, {\it i.e.} translation, time reversal,
reflection, and spin rotation
invariance, and the crystal point-group symmetries:
\begin{eqnarray}
F=  {\cal F}(\rho_{\vec k},{\bf S}_{\vec q}) 
+ {\cal F}(\rho_{{\vec k}^{\prime}}, {\bf S}_{{\vec q}^{\prime}})
+ {\cal F}_{int}(\rho_{\vec k}, {\bf S}_{\vec q},
\rho_{{\vec k}^{\prime}} , {\bf S}_{{\vec q}^{\prime}}) \nonumber 
\end{eqnarray}
\pagebreak
\begin{eqnarray}
{\cal F}= & & \frac {1} {2} r_{\rho} |\rho|^2 + U_{\rho} |\rho|^4  
\\ & & + \frac {1} {2} r_s |{\bf S}|^2 + U_s |{\bf S}|^4 
			+  U_x ({\bf S} \times {\bf S}^*) \cdot
			       ({\bf S} \times {\bf S}^*)       \nonumber
\\ & & + \lambda_1 [({\bf S} \cdot {\bf S}) \rho^* \ + \  c.c  ]    
 + 2 \lambda_2 |{\bf S}|^2 |\rho|^2 . \nonumber 
\end{eqnarray}
Note that it is not necessary to  include separate contributions from wave 
vectors $- \vec{q}$ and $- \vec{k}$ since the charge and spin densities are 
real, and hence $\rho_{- \vec{k}} = \rho^*_{\vec{k}}$ and
${\bf S}_{- \vec{q}} = {\bf S}^*_{\vec{q}}$. Also, there is no separate 
contribution of the form $|{\bf S} \cdot {\bf S}|^2$ because it may be 
written as a linear combination of the other quartic terms.
${\cal F}_{int}$, which is entirely quartic, has cross terms coupling
the order at $(\vec q,\vec k)$ and $(\vec q^{\prime},\vec k^{\prime})$ (
{\it e.g.} $V_1 |{\bf S}|^2||{\bf S}^{\prime}|^2$). 
For simplicity, and
because it is the case of experimental interest, it will be assumed that
the interactions in ${\cal F}_{int}$ are uniformly repulsive, so that
unidirectional order is favored ({\it i.e.} spin order will occur at $\vec q$
or $\vec q^{\prime}$, but not both).
For example, we do not allow ${\cal F}_{int}$ to contain terms that
favor checkboard order \cite{ps-stripes} as an alternative to stripe order. 

The third order term ($\lambda_1$) coupling spin and charge is allowed if
and only if
\begin{equation}
\vec k = 2\vec q .
\label{eq:k}
\end{equation}
As discussed previously\cite{topo}, this relation is the
generic reason for the ``topological'' character of the dopant-induced 
structure, as it implies that the period of the spin modulation is
twice the period of the charge modulation or, in other words, a periodic 
array of hole lines induces an array of antiphase spin domains. 

The free energy in Eq. (1) does not contain umklapp terms, in which the
sum of wave vectors is equal to a reciprocal lattice vector $\vec G$, which
become important in the neighborhood of a commensurability. Assuming that 
Eq. (2) is satisfied, the possible umklapp contributions up to fourth order are 
${\bf S}\cdot {\bf S}$ (when $2 \vec q = \vec G$);  $\rho^2$, 
${\bf S}\cdot {\bf S} \rho$, and $({\bf S}\cdot {\bf S})^2$ (all when
$4 \vec q = G$); $\rho^3$ (when 
$6 \vec q = G$) and $\rho^4$ (when $8 \vec q = G$).
Higher order terms give (weaker) commensurabilities at smaller wave
vectors. As usual \cite{BE}, the system will display
commensurate regions separated by soliton ``discommensurations'' when the
wave vector is close to a commensurate value.

From now on we will drop the subscripts on the order parameters, since each
order parameter has a single wave vector, as specified above. Also the 
normalization of the order parameters will be  chosen so that 
$U_{\rho} = 1 = U_s$.

{\bf  Nature of the ordered phases:}
With only two wave vectors ($\pm \vec q$), the spin order must be either
collinear or coplanar, since  a full three dimensional spin texture requires at 
least three ordering vectors.  
We consider the collinear and the non-collinear cases separately.

When the spin order is collinear, the axes and origin of 
coordinates may be chosen so that ${\bf S} = |S| {\bf e_1}$
and $\rho=|\rho|e^{i\theta}$, where $\theta$
determines the relative phase of the charge and spin density waves.  
In real space this means that the charge and spin density are
\begin{eqnarray}
&&\rho (\vec r)-\bar \rho = 2|\rho|\cos(2 \vec{q}\cdot \vec r - \theta), 
\nonumber \\ 
&& {\bf S}(\vec r) e^{i\vec Q \cdot \vec r} = 
2|S| \bf e_1 \cos(\vec{q}\cdot \vec r).
\end{eqnarray}
It is easily seen that
the free energy is minimized with $\theta=\pi$ for $\lambda_1>0$ and 
$\theta=0$ for $\lambda_1 <0$.
Clearly, for $\theta = \pi$, the charge density is peaked on the spin domain
walls where the magnitude of the spin order is zero. Since, in all microscopic 
models studied to date, doping tends
to depress magnetic order, we expect on general grounds that $\lambda_1 >0$.

Thus, for collinear spins, the Landau free energy can be expressed in terms of
the magnitudes of the order parameters as
\begin{eqnarray}
F_{linear} (|\rho|,|S|) &=&                                     
       \frac {1} {2} r_s |S|^2 +  |S|^4                      
\\ &+& \frac {1} {2} r_{\rho} |\rho|^2 +  |\rho|^4      \nonumber
\\ &-& 2 |\lambda_1| |S|^2 |\rho|                                         
    + 2 \lambda_2 |S|^2 |\rho|^2 .                                       \nonumber
\end{eqnarray}

For  a coplanar spiral phase, the origin of coordinates and axis of 
quantization may be chosen so that 
${\bf S}=|S| [\cos(\alpha) {\bf e}_1 \pm i \sin(\alpha) {\bf e}_2]$,
which corresponds to a real-space spin density 
\begin{eqnarray}                             
{\bf S}(\vec r) && e^{i\vec Q \cdot \vec r} =  \\
&& 2|S| [\cos(\alpha)\cos(\vec{q}\cdot \vec r){\bf e}_1 \pm
\sin(\alpha)\sin(\vec{q}\cdot \vec r){\bf e}_2]. \nonumber
\end{eqnarray}
Clearly, $\tan{\alpha}$ determines the eccentricity of the elliptical spiral,
and $\alpha=\pi/4$ corresponds to an ideal spiral, in which the magnitude of
the magnetic order is a constant.
The minimization of the free energy (with the assumption that $|S| \ne 0$)  
with respect to $\alpha$ (in the range $0 < \alpha \le \pi/4$)
and $\theta$ can be carried out straightforwardly.  If
$|\lambda_1||\rho|/U_x|S|^2 \ge 1$, the result is $\alpha=0$, or in other words
the collinear state is recovered.  For $|\lambda_1||\rho|/U_x|S|^2 <1$,
the free energy is minimized for
\begin{equation}
|S|^2 \cos(2\alpha) = |S_1|^2-|S_2|^2=|\lambda_1||\rho|/U_x
\label{eq:alpha}
\end{equation}
and as a function of 
$|\rho|$ and $|S|$  the free energy of the spiral state is
\begin{eqnarray}
F_{spiral} (|\rho|,|S|) &=&                                     
       \frac {1} {2} r_s |S|^2  + (1- U_x ) |S|^4              \nonumber
\\ &+& \frac {1} {2} (r_{\rho} - \frac {2 \lambda_1^2} {U_x}) |\rho|^2 
    +  |\rho|^4                                         \nonumber
\\ &+& 2 \lambda_2 |S|^2 |\rho|^2 .
\end{eqnarray}
The spiral phase is limited by the constraint $|\cos(2\alpha)| \le 1$,
which for the simple case of $\lambda_2 = 0$ is satisfied only for

\begin{eqnarray}
r_s^2 \left[ \frac {U_x} {2 |\lambda_1| (1-U_x)} \right]^2 +
\left( r_{\rho} - \frac {2 \lambda_1 ^2} {U_x} \right) \ge 0
\end{eqnarray}

\begin{figure}
\begin{center}
\leavevmode
\epsfxsize=3in \epsfbox{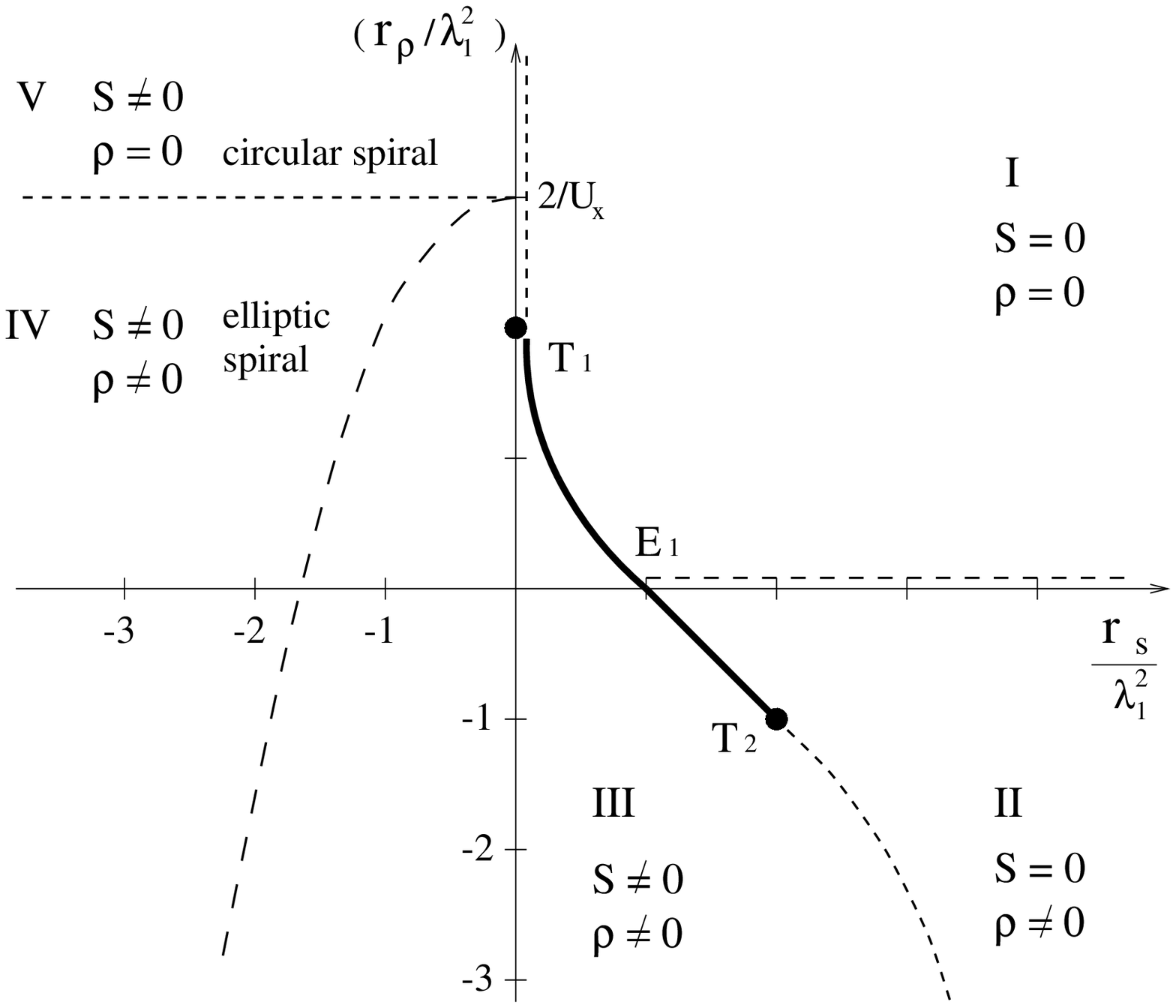}
\end{center}
\caption[Theoretical Phase Diagram]{\label{Phase Diagram:}}
\end{figure}

{\bf The phase diagram}  shown in Fig. 1 was derived by minimizing the free
energy in Eq. (1) with respect to $\rho$ and ${\bf S}$ where, to be
concrete, we have shown the locations of the phase boundaries for the
case $\lambda_2=0$ and $0 < U_x < 1$.
It exhibits five distinct phases:  {\bf  I)} a disordered
phase, with $|\rho| = 0$ and $|S| = 0$;  
{\bf  II)} a charge ordered phase with $|\rho| \neq 0$ and $|S| = 0$; 
{\bf  III)} a collinear stripe 
ordered phase with $|\rho| \neq 0$ and $|S| \neq 0$; 
{\bf  IV)} an elliptical-spiral phase, with
$|\rho|\neq 0$ and $|S|\neq 0$ and $0 < \alpha < \pi/4$ as
determined from Eq. (\ref{eq:alpha}). 
{\bf  V)} a circular spiral phase with
spin order but no accompanying charge order.  For $U_x \le 0$, the
two spiral phases are eliminated from the phase diagram, but the
remaining phase boundaries are unchanged.

Let us briefly sketch the analysis that leads to this phase diagram:

{\it  The Spin Driven Transition to a Collinear Phase:}
For $r_{\rho} > 0$, minimizing $F_{linear}$ with respect to $\rho$ gives
\begin{equation}  
|\rho| = {2 \lambda_1 |S|^2 \over r_{\rho}} + {\cal O} (|S|^4),
\end{equation}
and, on eliminating $\rho$ to obtain an
effective free energy for the spins,
\begin{equation}
F^{eff}_{linear} (|S|) =  \frac {1} {2} r_s |S|^2 +
(1- \frac {2 \lambda_1^2} {r_{\rho}}) |S|^4 +{\cal O}(|S|^6).    
\end{equation}
Clearly, for ${r_{\rho}} > {2 \lambda_1^2} $, 
there is a second order
transition as a function of $r_s$ from the disordered phase for
$r_s>0$ to the collinear stripe phase for $r_s<0$. 
This transition is spin driven; if we assume that $r_s \propto (T-T_c)$,
we find the usual mean-field exponent, 
$|S| \sim (T_c - T)^{1/2}$, while the charge modulation, which
is parasitic to the spin order, grows more
slowly, as $|\rho| \sim |S|^2 \sim (T_c - T)$. 
The second order line along $r_s = 0$ ends at a tricritical point, 
denoted by T$_1$ in Fig. 1, 
where $r_s=0$ and $r_{\rho}=2\lambda_1^2$, so that the coefficients
of both the $|S|^2$ and $|S|^4$ terms in $F^{eff} (|S|)$  vanish.

{\it  The Transition Driven by Spin-Charge Coupling:}
Below the tricritical point, where ${r_{\rho}} < 2{\lambda_1^2} $, 
the transition becomes first order, and moves into the 
quadrant in which both $r_s > 0$ and $r_{\rho} > 0$; 
here, the transition is driven by the coupling between spin 
and charge. The precise shape of the first order line depends, in general, on
$\lambda_2$ and on the values of higher order
terms neglected in our truncated form of the Landau free energy
in Eq. (1);  the general topology and structure of the phase
diagram shown in Fig. 1 are, however,  unchanged by these higher order terms.

{\it  The Charge Driven Transitions:}
For $r_s$ positive and sufficiently large, there
is a second order line along the $r_{\rho}=0$ axis
separating the disordered and charge-ordered
phases.  This line
ends at a critical end point, $E_1$ in the figure, where the first-order
line discussed previously crosses the axis.  
In this range of $r_s$, there is a second transition
at negative $r_{\rho}$
from the charge-ordered phase to the collinear stripe phase, at which
the spin-charge coupling finally causes the spin density to order as
well. Again, one can analyze this transition by first minimizing the
free energy with respect to $\rho$, and then analyzing $F^{eff}$
as a function of $S$. 
As a result, there is a second tricritical point, $T_2$, at which
this transition changes from first order (an extention of the
previously discussed first order line) to second order.  In either
case, the spin order, enhances the charge order. 
In this region of the phase diagram, higher order terms in the
Landau free energy can have quantatitive ( but, we believe, not
qualitative) effects on the results.  However, to be concrete,
it is useful to display analytic results
obtained with $\lambda_2$ and all higher order terms neglected.
Specifically, in this case, 
\begin{equation}  
\rho = \rho_0 + [\lambda_1  / |r_{\rho}|] |S|^2
+{\cal O} (|S|^4),
\end{equation}
where $\rho_0 = \sqrt{| r_{\rho}|}/2$ is the value of $\rho$ 
in the charge-ordered phase, $T_2$ is the point $r_s=2\lambda_1^2$
 and $r_{\rho}=-\lambda_1^2$, and
the second order line is given by the expression
$r_s=2\lambda_1\sqrt{|r_{\rho}|}$. The first order
line, which is given by the expression  $r_s=\lambda_1^2 + |r_{\rho}|$,
can be located straightforwardly once it is realized that on approaching
this line form the stripe-ordered side, $\rho=\lambda_1/2$ and
$|S|^2=(1/4) (2\lambda_1^2-r_s)$.

{\it The Transition to the Spiral Phases:}
For $U_x < 0$, it is easy to see that a collinear phase always has lower
free energy than any competitive spiral phase.  For $U_x >1$, the
Landau free energy as written is unbounded below, so higher order terms
must be included in any analysis.  
However, for $r_s <0$ and $0 < U_x < 1$,
there is the possiblility of a 
spiral phase. 

Therefore, to complete the phase diagram for $r_s <0$ and $0 < U_x < 1$,
one must compute the minimal value of $F_{spiral}$ (subject to the
constraint $|\cos(2\alpha)|=|\lambda_1||\rho|/U_x|S^2| \le 1$) and
compare it with the minimal value of $F_{linear}$.
For simplicity, we first consider the case $\lambda_2=0$.
It is easy then to see that for $r_{\rho} / \lambda_1^2 - 2/U_x > 0 $,
$F_{spiral}$ is minimized by $\rho=0$ and that
$F_{spiral}< F_{linear}$;  this is region V of the phase diagram, the
circular spiral-spin phase. 
At the point where $r_{\rho} / \lambda_1^2 - 2/U_x $ changes sign, 
it is clear from Eqs. (6) and (7) that there is a second order transition
(from region-V to region-IV ) marked by the onset of both charge order,
and an elliptical eccentricity to the spin spiral.
As $r_{\rho}$ decreases, the spiral eccentricity gradually increases 
until it achieves its maximum value (linear polarization).
The line $\cos(2\alpha)=1$ thus determines the phase boundary
between regions IV and III of the phase diagram;  for $\lambda_2=0$,
this line is simply the parabola determined by Eq. (8). 
The main effect of $\lambda_2 > 0$ (in this region of the phase diagram) 
is to expand the region of the circular spiral
phase at the expense of the elliptical spiral phase. 
For $\lambda_2 > \sqrt{ (1 - U_x)}$ the elliptical spiral phase is
completely eliminated.  Conversely, of course, $\lambda_2 <0$ tends to
stabilize the elliptical spiral phase.

{\bf Further Theoretical Considerations:}  

{\it  Effect of Higher Harmonics:}  So far we have considered only ordering
at the fundamental wave vector, although of course higher harmonics are   
induced below $T_c$. It is important to verify that
these harmonics do not affect the 
stability of the
various phases.  Slightly below $T_c$, in a spin-ordered phase, 
the effective free energy to order $|S_{n\vec q}|^2$ for the
$n^{th}$ harmonic is of the form
\begin{equation}
{\cal F}_n = r_n |{\bf S}_{n\vec{q}}|^2 + [{\bf A}_n\cdot {\bf S}_{n\vec{q}} + c.c.]
\end{equation}
where $r_n > 0$ and
${\bf A}_n$ is a function of ${\bf S}_{\vec q}$ and $\rho_{\vec k}$,
if there is charge order as well.  For a collinear phase, by rotational
invariance, ${\bf A}_n \propto {\bf S}$, so the induced higher harmonics
are always parallel to the fundamental.  Similarly, for a spiral phase with 
unbroken time-reversal symmetry, it is straightforward to see that 
${\bf A}_n$ must lie in the ordering plane, so that the planar character of
this phase is unaffected by higher harmonics.  However, if time reversal
symmetry is broken, then a contribution to ${\bf A}_0 \propto {\bf S}_{\vec q}
\times {\bf S}_{-\vec q}$ is allowed and the planar phase is unstable to the
formation of a three-dimensional spiral.  

{\it Goldstone Modes and Fluctuation Effects in Collinear Phases:} 
Landau theory is, of course, mean-field theory, so it is important to
address the effects of fluctuations about the mean-field state.  In the 
absence of commensurability effects, there is a Goldstone mode
which reflects the broken translational symmetry associated with finite
$\vec q$ ordering;  commensurability effects, if relevant,
will produce a gap in this
mode, which will be smaller the higher the order
of the commensurability.  In any of the
collinear spin-ordered phases, there are two Goldstone modes 
which reflect the broken spin-rotational symmetry.  Any uniaxial (Ising) 
anisotropy would produce a gap in these modes.

Thermal fluctuations of these
low lying modes may not dramatically
alter the phase diagram in three dimensions, but in quasi-two
dimensional systems, such as {\LNSCO}, they
always destroy the  long-range order, unless terms that break the symmetry
({\it e.g.} Ising anisotropy and umklapp scattering ) or inter-plane
couplings ({\it i.e.} three dimensional effects) are included in the analysis.
However, in many cases, this observation is accademic.  For instance,
the correlation length of the two-dimensional spin 1/2 Heisenberg model is 
roughly
proportional to $\exp (J/T)$ at low temperatures \cite{xi} (where $J$ is the 
exchange integral and $a$ is the lattice constant). Thus the magnetic 
correlation length can exceed the size of the sample at temperatures of 
interest and, for all practical purposes, the state
of the system is indistinguishable from long-range order.  The effects of
disorder are also potentially dramatic in two dimensions where the density
wave order will generally break up into Lee-Rice domains\cite{leerice}, but,
again, if the disorder is sufficiently weak, this may be of largely accademic
interest.

{\it  Goldstone Modes of the Spiral Phase:}  The elliptical spiral phase
has, in addition to the one Goldstone mode associated with broken
translational symmetry, and the two Goldstone modes associated
with rotations of the principal axis of the elipse, an
additional Goldstone mode associated with rotations about the principal axis.  
Ising anisotropy will open a gap in two of the spin-related Goldstone modes, 
but will leave the third one gapless;  it requires XYZ anisotropy
to fully gap the spin-related Goldstone modes.  Thus, the presence of a third
Goldstone mode, in particular one with an anomalously small gap, can be
used as a diagnostic for the presence of elliptical spin order.

In the circular spiral phase, rotations in the plane of the spiral are
equivalent to a translation, so the number of Goldstone modes is the
same as in the collinear phase.  However, the mode related to translational
invariance has a spin component, and thus is relatively insensitive to
commensurability effects and to the effects of disorder. 

{\bf Relation to experimental results:}
Several features of the experiments on the cuprates and nickelates may
be discussed in terms of the Landau theory analysis. Both 
{\LNSCO} \cite{NdCuO} and doped nickelates \cite{NiO} show an onset of charge 
order prior to spin order.  {\LNSCO} undergoes two continuous transitions 
(onset of charge order and later spin order), and therefore, can be 
associated with a path on the phase diagram that goes from phase-I to phase-II, 
and then to phase-III through second order transitions. 
Several {\LNiO} samples \cite{NiO} clearly show a 
first order transition associated with the onset of spin order in the
transition from phase-II ($|\rho| \neq 0$ , $|S| = 0$) to the full stripe
order, phase-III. Thus, these experiments can be associated with a path on
the phase diagram that goes through the first order transition line between
point $E_1$ and the tricritical point $T_2$. In both cases, it is clear that
the transitions are charge driven rather than spin driven.
Note that in this region of the phase diagram there is very little prospect 
of having a third phase transition to a spiral state, unless it occurs at 
much lower temperatures.

The above discussion has said nothing about the particular value of the wave 
vector $k=2q$. In the Landau theory, the value of $q$ can only depend on the 
(non-specific) $q$-dependence of the coefficients of the various terms in the 
free energy, or on commensurability. Experimentally, in the cuprates
\cite{NdCuO,CuO}, the 
ordered stripes in {\LNSCO} and the fluctuating stripes in {\LSCO} have 
$q \approx 
x$ for $x < 1/8$, which corresponds to one hole per two Cu along the 
stripe. For $x > 1/8$, $q \approx  
1/8$ in {\LSCO} and somewhat larger in
{\LNSCO}.   (Here, $q$ is measured in units of $2\pi/a$.)
This suggests that the value of
$q$ is influenced by the internal structure of the stripe for $x < 1/8$,
and by the $\rho^4$ umklapp term for $x > 1/8$. 
Furthermore, the temperature dependence of $q$ in the ordered phase of
{\LNSCO} is not strong, which suggests that the amplitude of the stripe is 
well established at the ordering temperature, and that the transition
is produced by phase ordering, as expected for quasi two-dimensional systems.
In all respects, the value of the stripe wave vector $q$ is determined by 
charge dynamics, rather than spin dynamics.

The conclusion of this analysis that stripes are charge driven rather than 
spin driven supports the idea that the driving force is 
Coulomb-frustrated phase separation \cite{ps,ps-stripes} driven by the hole 
dynamics. In order to minimize their kinetic energy, the holes attempt to
separate into hole-rich regions and regions with significant local
antiferromagnetic correlations. However phase separation is frustrated by 
the long-range Coulomb interaction between the holes, and the compromise between 
these two forces is the (charge-driven) stripe phase.
The antiphase ordering of the spin domains is a consequence of transverse
stripe fluctuations; it ensures that, on average, adjacent spins are 
antiparallel, whatever the location of the stripe.

The alternative mechanism for stripe formation is a Fermi surface instability
\cite{HF-stripes} (due to nesting), in which the spins form antiphase domains that are 
stabilized by holes bound to the domain walls. In this scenario, spin and
charge order together or charge stripe order follows after spin order,
contrary to experiment. 
These microscopic theories also predicted the possibility of spiral phases
\cite{shraiman}
(though not that they must be elliptical). We conclude that theories based
on a Fermi surface instability may be relevant in the spin-driven region
of the phase diagram $(r_{\rho} > 0)$ but not for any of the materials of
interest, which all show a charge driven sequence of transitions.

Indeed there are more general reasons to believe that the
ordered phases in the high-T$_c$ cuprates should not arise from a Fermi
surface instability. It has been argued \cite{noqp} that these materials
belong to a class of ``bad metals'', in which there are no quasiparicles,
and therefore no Fermi surface or Fermi surface instabilities.

{\bf Acknowledgements:} We thank John Tranquada for numerous discussions of
his experimental results on the cuprates and nickelates, and especially the
implications of the Landau theory.
This work was supported in part by the National Science Foundation grant 
number DMR93-12606 (SK \& OZ) at UCLA.  Work at
Brookhaven (VE) was supported by the Division of Materials Science,
U. S. Department of Energy under contract No. DE-AC02-76CH00016.

\end{document}